\documentclass[reprint,a4paper,aps,pra,amsmath,amsfonts,showpacs,floatfix,nofootinbib]{revtex4-1}
\usepackage{times}
\usepackage{amsmath}
\usepackage{times}
\usepackage{dcolumn}                 % Aligning table columns
\usepackage{graphicx}
\usepackage{xcolor}

\newcommand*{\cent}[1]{\multicolumn{1}{c}{$#1$}}
\newcolumntype{x}[1]{D{.}{.}{#1}}

\begin{document}
\title{Hyperfine structure in the HD molecule}

\author{Jacek Komasa}
\affiliation{Faculty of Chemistry, Adam Mickiewicz University, Uniwersytetu Pozna{\'n}skiego 8, 61-614 Pozna{\'n}, Poland}

\author{Mariusz Puchalski}
\affiliation{Faculty of Chemistry, Adam Mickiewicz University, Uniwersytetu Pozna{\'n}skiego 8, 61-614 Pozna{\'n}, Poland}

\author{Krzysztof Pachucki}
\affiliation{Faculty of Physics, University of Warsaw, Pasteura 5, 02-093 Warsaw, Poland}

\begin{abstract}
We investigate interactions between the proton spin, the deuteron spin, and
the orbital angular momentum in the electronic ground state of the HD molecule. 
These interactions lead to hyperfine splittings of molecular energy levels. 
Our numerical results for the first rotational level agree well with the currently most accurate
measurement performed by Ramsey {\em et al.} in the 1950s. 
Knowledge of the hyperfine structure of other levels is necessary
for the accurate determination of rovibrational transition energies in spectroscopic measurements.
We present theoretical predictions and share the numerical code used to perform numerical calculations.
This work sets the ground for high precision spectroscopic tests of hyperfine interactions in molecular systems.
In particular we determine the value of the deuteron quadrupole moment $Q = 0.2856(2)$ fm$^2$ 
and give outlook for improving its accuracy by three orders of magnitude.
\end{abstract}
\maketitle

\section{Introduction}
Current theoretical predictions for the hyperfine structure (hfs) in simple molecules
are far less accurate than the experimental values obtained in original microwave measurements 
by Ramsey {\em et al.}~\cite{Quinn:58,Code:71} half a century ago. 
For HD these measurements were performed, regrettably, only for the lowest rotational level.
On the other hand, theoretically predicted hyperfine structure (hfs) of many other 
molecular levels is necessary for a contemporary determination of transition frequencies,
due to the complexity of the line shape.
For instance, several recent measurements \cite{Fasci:18,Cozijn:18,Tao:18}
of a specific transition frequency in HD significantly disagree with each other, 
presumably because of a different line shape theory that has been employed. 
Therefore, the knowledge of the hyperfine splitting and corresponding individual 
transition rates is of crucial importance in a correct interpretation of precision molecular spectroscopy.

The first step in this direction is the recent work of Dupr\'e \cite{Dupre:20}, which presents
results for low lying vibrational levels with the rotational number $J=1$, however without any uncertainties.
In this work we present a systematic derivation and numerical calculation of leading hyperfine interactions
for all molecular levels in the HD molecule, including individual hfs transition rates between them,
within the Born-Oppenheimer (BO) approximation. Due to this approximation
our results will have about $10^{-3}$ relative accuracy, which nevertheless is sufficient for
the current experimental precision of rovibrational transitions. Moreover, we present hyperfine splittings
for an arbitrary rovibrational level of HD in terms of a freely available computer code \cite{H2Spectre}.
As well as application in molecular spectroscopy, our results could also be useful in precision 
tests of hyperfine interactions in the HD molecule, provided the theory for relativistic 
and quantum electrodynamic corrections is developed.

Considering theory for the leading molecular hyperfine structure,
there are three angular momenta in the ground electronic state of the HD molecule:
the proton spin $\vec{I}_p$, that of the deuteron $\vec{I}_d$, and the rotational angular momentum $\vec{J}$. 
All of them interact with each other, and using the Ramsey notation (e.g. \cite{Ramsey:57}), 
the effective Hamiltonian describing these interactions reads
\begin{widetext}
\begin{align}
  H_{\rm hfs} =&\ -c_p\,\vec I_p\cdot\vec J - c_d\,\vec I_d\cdot\vec J
%\nonumber \\ &
+ \frac{5\,d_1}{(2\,J-1)(2\,J+3)}\,\biggl[
      \frac{3}{2}\,(\vec I_p\cdot\vec J)\, (\vec I_d\cdot\vec J)
      +\frac{3}{2}\,(\vec I_d\cdot\vec J)\, (\vec I_p\cdot\vec J)
      -(\vec I_p\cdot\vec I_d)\,\vec{J}^{\;2}\biggr]
\nonumber \\ &
+ \frac{5\,d_2}{(2\,J-1)(2\,J+3)}\,\biggl[
      3\,(\vec I_d\cdot\vec J)^2
      +\frac{3}{2}\,(\vec I_d\cdot\vec J)
      -\vec I_d^{\;2}\,\vec J^{\;2}\biggr]\,.
\label{01}
\end{align}
\end{widetext}
One observes the lack of a separate $\vec I_p\cdot \vec I_d$ coupling.
The direct scalar nuclear spin-spin interaction vanishes, while the electron-mediated
nuclear spin-spin interaction is of higher order in the fine structure constant $\alpha$. 
Namely, it is $\alpha^2$-times smaller than the above tensor interactions and
therefore is neglected here as are all the other $\alpha^2$ corrections.

The above coefficients $c_p$, $c_d$, $d_1$, and $d_2$ are related, respectively, 
to the interactions between: the proton spin and molecular rotation, 
the deuteron spin and rotation, the proton and deuteron spins, and the electric
quadrupole moment of the deuteron with the electric field gradient. 
All of these constants depend on the molecular level, identified by the vibrational $v$ 
and the rotational $J$ quantum numbers.

In the following Sections, we present a short derivation of all the constants, followed by their
numerical calculation as functions of the internuclear distance $R$. Their values for a particular 
molecular state are obtained by averaging with the nuclear wave function $\chi_{v,J}$
\begin{align}
  H_{\rm hfs}(v,J) = \langle v,J|H_{\rm hfs}(R)|v,J\rangle\,. \label{Ehfsaver}
\end{align}
This is an approximate treatment that relies on the BO approximation, but it is a good starting point
for future more accurate nonadiabatic calculations.
Ramsey, in his monograph on molecular beams \cite{Ramsey:56}, presented a basic theory of nuclear and rotational
magnetic moment coupling. Here, we present a concise and rigorous
derivation of all molecular hfs interactions, which can also be a basis for the derivation of relativistic 
as well as QED corrections.

\section{Nuclear spin-rotation constants $c_p$ and $c_d$}
The following derivation of spin-rotation constant is based on our former work 
\cite{Pachucki:10}, from which we adopt the notation. 
Let us consider a Hamiltonian for a particle with charge $e$, mass $M$, spin $I$,
and a gyromagnetic factor $g$, interacting with the electromagnetic field $(\hbar = c = 1)$,
\begin{align}
  H &=\frac{\vec\pi^2}{2\,M} + e\,A^0 -\frac{e\,g}{2\,M}\,\vec I\cdot\vec B \nonumber\\
    &\quad -\frac{e\,(g-1)}{4\,M^2}\,\vec I\cdot(\vec E\times\vec\pi-\vec\pi\times\vec E)\,,
\label{03}
\end{align}
where $\vec\pi = \vec p -e\,\vec A$.
If this particle is the proton or deuteron, then $e$ coincides with the elementary charge.
The gyromagnetic factors
\begin{align}
  g_p =&\ \frac{\mu_p}{\mu_N\,I_p} =5.585\,695\ldots \\
  g_d =&\ \frac{\mu_d}{\mu_N\,I_d}\,\frac{m_d}{m_p} = 1.714\,025\ldots
\end{align}
are related to the magnetic moment of the proton $\mu_p = 2.792\,847\,344\,63(82)\,\mu_N$ and the deuteron $\mu_d = 0.857\,438\,2338(22)\,\mu_N$, respectively \cite{CODATA:18}. The $g$-factor is a dimensionless quantity which is more convenient to use in formulas than the magnetic moment. If the electromagnetic field comes from the other nucleus or from the electron $(g=2, M=m)$, it is of the form
\begin{align}
  \vec E =&\ -\frac{e}{4\,\pi}\,\frac{\vec r}{r^3}\,,  \\
  A^i =&\ \frac{e}{4\,\pi}\biggl[\frac{1}{2\,r}\biggl(\delta^{ij}+\frac{r^i\,r^j}{r^2}\biggr)\,p^j-\frac{g}{2\,M}\,\vec I\times\frac{\vec r}{r^3}\biggr]\,,
\end{align}
where $i,j$ are Cartesian indeces.
Inserting the above formulas to Eq. (\ref{03}) one obtains the general spin-orbit Hamiltonian
\begin{align}
\delta H =& \sum_{\alpha,\beta}\frac{e_\alpha\,e_\beta}{4\,\pi}\,\frac{1}{2\,r_{\alpha\beta}^3}
\biggl[\frac{g_\alpha}{m_\alpha\,m_\beta}\,\vec I_\alpha\cdot\vec r_{\alpha\beta}\times\vec p_\beta
\nonumber \\ &\ 
  -\frac{(g_\alpha-1)}{m_\alpha^2}\,\vec I_\alpha\cdot\vec r_{\alpha\beta}\times \vec p_\alpha \biggr],
\label{EgsoH}
\end{align}
where the indices $\alpha$ and $\beta$ go over both electrons and nuclei. 
In particular, the coupling of the nuclear spin $\vec I = \vec I_A$ 
to the molecular rotation is
\begin{align}
  \delta_A H =&\ \sum_{b}\frac{e_A\,e}{4\,\pi}\,\frac{\vec I}{2\,r_{A b}^3}\!
  \biggl[\frac{g_A}{m_A\,m}\,\vec r_{A b}\times\vec p_b-
       \frac{(g_A-1)}{m_A^2}\,\vec r_{A b}\times \vec p_A \biggr]
\nonumber \\ &\hspace{-6ex}
+\frac{e_A\,e_B}{4\,\pi}\,\frac{\vec I}{2\,r_{A B}^3}\cdot
\biggl[\frac{g_A}{m_A\,m_B}\,\vec r_{A B}\times\vec p_B-
       \frac{(g_A-1)}{m_A^2}\,\vec r_{A B}\times \vec p_A \biggr].
\end{align}
For convenience, following Ref.~\onlinecite{Pachucki:10}, we chose the reference frame centered at
the considered nucleus $A$ and introduced the notation $\vec R = \vec r_{AB}$, 
$\vec P = -\imath\,\vec \nabla_{\!R}$, and $\vec x_b = \vec r_{bA}$.
For the $\Sigma^+_g$ electronic state considered here, $\delta_A H$ takes the form
\begin{align}
\delta_A H &= \vec Q_1\cdot\vec I + \vec Q_2\times\vec P\cdot\vec I,\\
\vec Q_1
&=-\sum_{b}\frac{e_A\,e}{4\,\pi}\,\frac{g_A}{2\,m\,m_A}\,\frac{\vec x_b\times \vec p_b}{x_b^3},\\
\vec Q_2 &= -\sum_{b}\frac{e_A\,e}{4\,\pi}\,\frac{1}{2\,m_A^2}\,\frac{\vec x_b}{x_b^3}\nonumber\\&\quad
-\frac{e_A\,e_B}{4\,\pi}\,\frac{1}{2\,m_A}\,
\biggl[\frac{g_A}{m_B}+\frac{(g_A-1)}{m_A} \biggr]\,\frac{\vec R}{R^3}, \label{EdHQ2}
\end{align} 
where we neglected terms of the higher order in the electron-nucleus mass ratio.
We make use of BO approximation, and the total wave function $\psi$ is represented as a product
\begin{align}
  \psi_{v,J,M} = \phi_{\rm el}(\vec x_a)\,\chi_{v,J}(R)\,Y_{J,M}(\vec n)
\end{align}
of the electronic wave function $\phi_{\rm el}(\vec x_a)$, the nuclear one $\chi_{v,J}(R)$,
and the spherical harmonic $Y_{J,M}(\vec n)$, where $\vec n = \vec R/R$.
The electronic wave function $\phi_{\rm el}$ for the ground $\Sigma^+$ state is a scalar function,
and thus depends only on interparticle distances.

The expectation value of $\langle\phi_{\rm el}|\vec Q_1|\phi_{\rm el}\rangle$
vanishes and the $\vec Q_1$ operator contributes only through the nonadiabatic matrix element \cite{Pachucki:10}
\begin{equation}
\langle \vec Q \rangle_{\rm el}^{(1)} =
-\frac{\vec R\times\vec P}{m_{\rm n}\,R^2}\
\left\langle\phi_{\rm el}\left|J^j_{\rm el}\,
\frac{1}{({\cal E}_{\rm el}-H_{\rm el})'}\,Q^j\right|\phi_{\rm el}\right\rangle.
\end{equation}
where $\vec J_{\rm el}$ is the electronic angular momentum operator, and $1/m_{\rm n} = 1/m_A + 1/m_B$,
so that the total spin-rotation constant $c_A$ can be inferred from
\begin{align}
-c_A\,\vec I\cdot\vec J &= -\frac{\vec I\cdot\vec J}{m_{\rm n}\,R^2}\,
\langle\phi_{\rm el}|\vec J_{\rm el}\,\frac{1}{({\cal E}_{\rm el}-H_{\rm el})'}\,\vec
Q_1|\phi_{\rm el}\rangle \nonumber\\&\quad
+ \langle\phi_{\rm el}|\vec Q_2|\phi_{\rm el}\rangle\times\vec P\cdot\vec I\,.
\end{align}
The expectation value of the first term in $\vec Q_2$ (Eq.~(\ref{EdHQ2})) can alternatively be expressed in
terms of a derivative of the BO energy, namely
\begin{eqnarray}
\left\langle\phi_{\rm el}\left|\sum_{b}\frac{e_A\,e}{4\,\pi}\,\frac{\vec x_b}{x_b^3}\right|\phi_{\rm el}\right\rangle
&=&
\vec n\,\biggl(\frac{\partial{\cal E}_{\rm el}}{\partial R}+\frac{e_A\,e_B}{4\,\pi}\,\frac{1}{R^2}\biggr),
\end{eqnarray}
and $c_A$ in atomic units $[e_X =-Z_X\,e, \alpha=e^2/(4\,\pi)]$ becomes
\begin{widetext}
\begin{eqnarray}
c_A &=&\alpha^2\biggl[\frac{1}{R^2}\,\frac{Z_A\,g_A}{2\,m_{\rm n}\,m_A}\,
\left\langle\phi_{\rm el}\left|\sum_a\vec x_a\times\vec p_a\,\frac{1}{({\cal E}_{\rm el}-H_{\rm el})}
\sum_{b} \frac{\vec x_b\times \vec p_b}{x_b^3}\right|\phi_{\rm el}\right\rangle
+ \frac{1}{R}\,\frac{1}{2\,m_A^2}\,\frac{\partial{\cal E}_{\rm el}}{\partial R}
+\frac{1}{R^3}\frac{Z_A\,Z_B\,g_A}{2\,m_{\rm n}\,m_A}\biggr].
\end{eqnarray}
In the particular case of the proton and the deuteron in the HD molecule, we arrive at
\begin{align}\label{Ecp}
  c_p =&\ \frac{\alpha^2}{2\,R^2\,m_{\rm n}\,m_p}\,\biggl[
    g_p\,\left\langle\phi_{\rm el}\left|\sum_a\vec x_a\times\vec p_a\,\frac{1}{({\cal E}_{\rm el}-H_{\rm el})}
    \sum_{b} \frac{\vec x_b\times \vec p_b}{x_b^3}\right|\phi_{\rm el}\right\rangle
    +\frac{g_p}{R} + \frac{m_{\rm n}}{m_p}\,R\,\frac{\partial{\cal E}_{\rm el}}{\partial R}\biggr], \\\label{Ecd}
  c_d =&\ \frac{\alpha^2}{2\,R^2\,m_{\rm n}\,m_d}\,\biggl[
    g_d\,\left\langle\phi_{\rm el}\left|\sum_a\vec x_a\times\vec p_a\,\frac{1}{({\cal E}_{\rm el}-H_{\rm el})}
    \sum_{b} \frac{\vec x_b\times \vec p_b}{x_b^3}\right|\phi_{\rm el}\right\rangle
    +\frac{g_d}{R} + \frac{m_{\rm n}}{m_d}\,R\,\frac{\partial{\cal E}_{\rm el}}{\partial R}\biggr].  
\end{align}
\end{widetext}
These formulas coincide with those derived originally in \cite{Reid:74}.

\section{Spin-spin constant $d_1$}
The nuclear spin-spin direct interaction comes from the 3rd term in Eq. (\ref{03}) and is of the form
\begin{align}
  \delta H = \frac{e_A\,e_B}{4\,\pi}\,\frac{g_A\,g_B}{4\,m_A\,m_B}\,\frac{I_A^i\,I_B^j}{R^3}\,\biggl(\delta^{ij}-3\,\frac{R^i\,R^j}{R^2}\biggr).
\end{align}
Using the fact that for the $\Sigma$ states of a diatomic molecule $\vec n\cdot\vec J |J,M_J\rangle=0$,
the matrix elements of the angular part (in parentheses) in states with definite angular momentum $J$
can be expressed in terms of this $J$ as
\begin{align}
&\biggl\langle J,M_J\left|\delta^{ij}-3\,\frac{R^i\,R^j}{R^2}\right|J,M'_J\biggr\rangle \nonumber \\
& = \biggl\langle J,M_J\left|\frac{3\,J^i\,J^j + 3\,J^j\,J^i -2\,\delta^{ij}
 \,\vec J^2}{(2\,J-1)\,(2\,J+3)}\right|J,M'_J\biggr\rangle.
\end{align}  
So, for such states, this interaction takes the form (in atomic units)
\begin{align}
  \delta H &= \alpha^2\,\frac{g_p\,g_d\,m_e^2}{4\,m_p\,m_d}\,\frac{1}{R^3}\times\\&\quad
  \frac{3\,(\vec I_p\cdot\vec J)\,(\vec I_d\cdot\vec J) + 3\,(\vec I_d\cdot\vec J)\,(\vec I_p\cdot\vec J)
      -2\,(\vec I_p\cdot\vec I_d)\,\vec J^2}{(2\,J-1)\,(2\,J+3)}\nonumber
\end{align}
and the $d_1$ constant is
\begin{align}\label{Ed1}
  d_1&=\alpha^2\,\frac{g_p\,g_d\,m_e^2}{10\,m_p\,m_d}\,\frac{1}{R^3} \,.
\end{align}

\section{Quadrupole constant $d_2$}
The interaction of a particle possessing the electric quadrupole moment with 
the gradient of the electric field is given by
\begin{align}
  \delta H =&\ -\frac{e}{6}\,Q^{ij}\,\partial_jE^i\,.
\end{align}
For a particle with a definite spin $I\geq 1$, the $Q^{ij}$, as a traceless and symmetric tensor, 
can be expressed in terms of a single scalar electric quadrupole moment $Q$ defined by 
\begin{align}
  Q^{ij} =&\ \frac{Q}{I\,(2I-1)}\,\biggl(\frac{3}{2}\,I^i\,I^j + \frac{3}{2} I^j\,I^i -\delta^{ij}\,\vec I^2\biggr).
\end{align}
This definition is such that $Q$ corresponds to the expectation value of $Q^{33}$ in a state
with the maximum value of $M_I$, namely
\begin{align}
  Q = \langle I,I|Q^{33}|I,I\rangle\,.
\end{align}
The electric field is produced by the other nucleus and all the electrons. Let us introduce
$q$, which is an averaged value of the gradient of the molecular electric field
\begin{align}
q &\equiv \frac{1}{3}\,\langle\phi_{\rm el}| e\,\partial_jE^i|\phi_{\rm el}\rangle 
 \,\biggl(\delta^{ij}-3\,\frac{R^i\,R^j}{R^2}\biggr) \nonumber\\ 
&= \bigg\langle\phi_{\rm el}\left| \frac{\partial^2 V}{\partial R_d^i\partial R_d^j}\,
  \biggl(\frac{R^i\,R^j}{R^2} - \frac{\delta^{ij}}{3}\biggr)\right|\phi_{\rm el}\bigg\rangle, \label{Eq}
\end{align}
where $V$ is the Coulomb interaction potential.
Then, the traceless part of the electric field gradient is
\begin{align}  
&\langle\phi_{\rm el}| e\,\partial_j E^i|\phi_{\rm el}\rangle
  =\frac{q}{2}\,\biggl(\delta^{ij}-3\,\frac{R^i\,R^j}{R^2}\biggr)\,
\end{align}
and in a state with  the definite angular momentum
\begin{align}
&\langle J,M_J,\phi_{\rm el}| e\,\partial_j E^i|\phi_{\rm el},J,M'_J\rangle\nonumber \\  
&=\frac{q}{2}\, \biggl\langle J,M_J\left|\frac{3\,J^i\,J^j + 3\,J^j\,J^i -2\,\delta^{ij}\,\vec J^2}{(2\,J-1)\,(2\,J+3)}\right|J,M'_J\biggr\rangle.
\end{align}
Finally, the interaction of the electric quadrupole moment of the nucleus with the gradient 
of the molecular electric field is
\begin{align}
\delta H &=-\frac{1}{6}\,
\frac{Q}{I\,(2I-1)}\,\biggl(\frac{3}{2}\,I^i\,I^j + \frac{3}{2} I^j\,I^i -\delta^{ij}\,\vec I^2\biggr)
\nonumber\\&\quad
 \times\frac{q}{2}\, \frac{3\,J^i\,J^j + 3\,J^j\,J^i -2\,\delta^{ij}\,\vec J^2}{(2\,J-1)\,(2\,J+3)}
 \nonumber \\ 
&=-Q\,q\,\frac{3\,(\vec I\cdot\vec J)^2 + \frac{3}{2}\,(\vec I\cdot\vec J)-\,\vec I^2\,\vec J^2}
  {2\,I\,(2I-1)\,(2\,J-1)\,(2\,J+3)}.
\end{align}  
The Ramsey constant $d_2$ is thus 
\begin{align}
  d_2 &= -\frac{Q\,q}{10} ,
  \intertext{which in atomic units reads}\label{Ed2}
  d_2&=-\alpha^2\,\frac{Q}{10\,\lambdabar^2}\,\left\langle\phi_{\rm el}\left|\frac{\partial^2 V}{\partial R_d^i\,\partial R_d^j}\,
  \biggl(\frac{R^i\,R^j}{R^2}-\frac{\delta^{ij}}{3}\biggr)\right|\phi_{\rm el}\right\rangle,
\end{align}
where $\lambdabar$ is the reduced Compton wavelength of an electron.

\section{Numerical calculations of hyperfine curves}

To evaluate the electronic matrix elements present in $c_p$ and $c_d$ (see Eqs.~(\ref{Ecp}) and (\ref{Ecd}))
we use explicitly correlated Gaussian (ECG) basis functions of $\Sigma^+$
\begin{equation}
  \phi_\Sigma = e^{-a_{1A}\,r_{1A}^2 -a_{1B}\,r_{1B}^2 -a_{2A}\,r_{2A}^2 -a_{2B}\,r_{2B}^2 - a_{12}\,r_{12}^2 }  \label{32}
\end{equation}
and $\Pi$ symmetry 
\begin{equation}
  \phi_{\Pi}^i = (\vec R \times \vec r_{1})^i\,\phi = \epsilon^{ijk} R^j r_1^k\, \phi_\Sigma \,. \label{33}
\end{equation}
256  basis functions of Eq.~(\ref{32}) were employed to represent the electronic wave function $\phi_{\rm el}$.
The same number of functions of Eq.~(\ref{33}) was used to form the internal basis set
of the resolvent $1/({\cal E}_{\rm el}-H_{\rm el})$.
Their nonlinear parameters were determined variationally in a global optimization process
independently at 44 internuclear distances. While the parameters of the $\phi_{\rm el}$ were determined
by minimizing the electronic energy, the nonlinear parameters of the internal basis
were optimized with respect to the functional
\begin{equation}
\langle\phi_{\rm el}|\sum_a\vec x_a\times\vec p_a\,\frac{1}{{\cal E}_{\rm el}-H_{\rm el}}
    \sum_{b} \vec x_b\times \vec p_b|\phi_{\rm el}\rangle\,.
\end{equation}
Thanks to the optimization of the $\phi_\Sigma$ and $\phi_\Pi$ functions the relative numerical accuracy 
(ca. $10^{-5}$) of the spin-rotation parameters is higher than the estimate of nonadiabatic corrections, 
and the use of only 256 basis functions was sufficient for this purpose.
Apart from the second-order matrix element, the spin-rotation parameters $c_p$ and $c_d$
require evaluation of the derivative of the BO energy with respect to the intermolecular distance, 
see Eqs.~(\ref{Ecp})-(\ref{Ecd}). This derivative can be found from the virial theorem
\begin{align}
\frac{\partial{\cal E}_{\rm el}}{\partial R}&=\frac{\langle V\rangle_\mathrm{el}-2\,\mathcal{E}_\mathrm{el}}{R}\,,
\end{align}
which enables calculations with high numerical precision.

The direct spin-spin interaction constant $d_1$ does not require evaluation of any electronic matrix elements 
and, for a given $R$, is fully determined by the well-known nuclear $g$-factors and the electron-nucleus mass ratios.

Considering the matrix element of the quadrupole constant $d_2$ in Eq.~(\ref{Ed2}),
we integrated it by parts to obtain a less singular form, 
\begin{align}
\biggl\langle\frac{\partial^2 V}{\partial R_d^i\,\partial R_d^j}\,\biggl(\frac{R^i\,R^j}{R^2}-\frac{\delta^{ij}}{3}\biggr)\biggr\rangle =
 \frac{2}{R^3} -\biggl(\frac{R^i\,R^j}{R^2}-\frac{\delta^{ij}}{3}\biggr)\nonumber\\
\times\int d^3r_1\,d^3r_2\biggl(
  \frac{1}{r_{1A}}\frac{\partial^2 (\phi^2_{\rm el})}{\partial r_{1}^i\,\partial r_{1}^j} +
   \frac{1}{r_{2A}}\frac{\partial^2 (\phi^2_{\rm el})}{\partial r_{2}^i\,\partial r_{2}^j}\biggr),
\end{align}
which is more convenient in calculations. The above expectation value was evaluated with 
$\phi_{\rm el}$ expanded in an ECG basis as large as 1024 terms, due to slow numerical convergence.
Table~\ref{Tqconv} supplies data which enable an analysis of this convergence at different
regions of the internuclear distance. This analysis reveals that, depending on the region,
4-6 significant digits are stable. Our numerical results are in good agreement with the results
published by Pavanello {\em et al.}~\cite{Pavanello:10} except for the shortest internuclear distances,
at which their values seem to be less accurate.
As a final result we take values from the 1024-term basis and note that the achieved numerical accuracy 
of the electric field gradient within the BO approximation is higher than the estimated 
contribution from the nonadiabatic effects.

In contrast to previously described magnetic interactions,
the electric quadrupole interaction constant $d_2$ depends on the electric quadrupole moment of the deuteron $Q$,
which is not well known from independent measurements. In fact, it is the old Ramsey measurement~\cite{Code:71},
which allows the most accurate determination of the deuteron quadrupole moment.
For this purpose we use the measurement for the $J=1$ level of the  D$_2$ molecule,
for which $d_2$ is found with the highest accuracy and the nonadiabatic effects are smaller
in comparison to the HD molecule. This determination of $Q$ is described in detail in the next Section,
however, we use this value here for evaluation of the $d_2$ curve.
The final numerical results for the spin-rotation ($c_p$ and $c_d$, Eqs.~(\ref{Ecp}) and ~(\ref{Ecd})), 
spin-spin ($d_1$, Eq.~(\ref{Ed1})), the electric field gradient ($q$, Eq.~(\ref{Eq})),
and quadrupole ($d_2$, Eq.~(\ref{Ed2})) constants for all
the internuclear distances are presented in Table~\ref{Tconstants}.
The conversion factor from energies in atomic units to frequencies in Hz is $2\,Ry\,c$, 
where $Ry$ is the Rydberg constant and $c$ is the speed of light in a vacuum.
For small $R$, all curves exhibit $R^{-3}$ dependence as they should, while
for large $R$ they vanish faster than $R^{-3}$. 

\begin{table}
\caption{Convergence of the electric field gradient $q$ defined in Eq.~(\ref{Eq}) 
with the growing basis set size $K$ at selected internuclear distances $R$ in comparison with 
the most accurate literature data (all data in atomic units).}
\label{Tqconv}
\begin{ruledtabular}
\begin{tabular}{r@{\extracolsep{\fill}}x{2.6}x{1.6}x{2.9}}
$K$&  \cent{R=0.4}  &  \cent{R=1.4}  & \cent{R=5.0} \\
  \hline\\[-1.5ex]
 128                    & 30.082\,378 & 0.338\,173 & -0.001\,890\,827 \\
 256                    & 30.082\,224 & 0.338\,084 & -0.001\,887\,565 \\
 512                    & 30.082\,195 & 0.338\,078 & -0.001\,888\,152 \\
1024                    & 30.082\,184 & 0.338\,073 & -0.001\,890\,408 \\[2ex]
\cite{Pavanello:10}$^a$ & 30.405\,155 & 0.338\,070 & -0.001\,890\,88  \\
\cite{Reid:73}$^b$      &             & 0.336\,30 \\
\end{tabular}
\end{ruledtabular}
\flushleft
$^a$ Pavanello {\em et al.} (2010) \cite{Pavanello:10}.\\
$^b$ Reid and Vaida (1973) \cite{Reid:73}.
\end{table}

\section{Hyperfine constants}
%
% Table from Aver_hyperfine.nb and Limit_Rto0.nb
\renewcommand*{\arraystretch}{1}
\begin{table} %\scriptsize
\caption{The hyperfine splitting parameters (in kHz) and electric field gradient $q$ (in a.u.) evaluated 
with ECG wave functions at different internuclear distances, $R$ (in a.u.). According to Eq.~(\ref{Ed1}), 
$d_1(R)\,R^3=49.7735$. The deuteron quadrupole moment used here is $Q=0.2856(2)\;{\rm fm}^2$, see Eq.~(\ref{EQD}).
The relative numerical uncertainty of $c_p$ and $c_d$ is about $10^{-5}$, while that of $q$ and $d_2$ is below  $10^{-4}$
with the exception of large distances i.e. $R>4$ a.u.  
}
\label{Tconstants}
\begin{ruledtabular}
\begin{tabular}{c@{\extracolsep{\fill}}x{4.2}x{3.5}x{2.5}x{5.5}}
$R$&  \cent{c_p(R)\, R^3}  &  \cent{c_d(R)\, R^3}  & \cent{q(R)\, R^3} & \cent{d_2(R)\, R^3} \\
  \hline\\[-1.5ex]
0.00 & 383.478 & 53.8392 & 2       & -134.306 \\
0.05 & 383.343 & 53.8227 & 1.99993 & -134.206 \\
0.10 & 382.562 & 53.7266 & 1.99910 & -134.150 \\
0.20 & 378.132 & 53.1835 & 1.99045 & -133.570 \\
0.30 & 370.082 & 52.2006 & 1.96675 & -131.979 \\
0.40 & 359.411 & 50.9042 & 1.92526 & -129.195 \\
0.50 & 347.140 & 49.4211 & 1.86651 & -125.253 \\
0.60 & 334.055 & 47.8480 & 1.79257 & -120.291 \\
0.80 & 307.471 & 44.6783 & 1.60929 & -107.992 \\
1.00 & 282.174 & 41.6944 & 1.39463 & -93.5873 \\
1.10 & 270.345 & 40.3096 & 1.28044 & -85.9243 \\
1.20 & 259.127 & 39.0020 & 1.16374 & -78.0931 \\
1.30 & 248.527 & 37.7708 & 1.04579 & -70.1780 \\
1.40 & 238.531 & 36.6121 & 0.927672 & -62.2517 \\
1.50 & 229.121 & 35.5221 & 0.810308 & -54.3760 \\
1.60 & 220.252 & 34.4935 & 0.694516 & -46.6057 \\
1.70 & 211.885 & 33.5194 & 0.581002 & -38.9883 \\
1.80 & 203.974 & 32.5925 & 0.470417 & -31.5675 \\
1.90 & 196.458 & 31.7028 & 0.363340 & -24.3820 \\
2.00 & 189.296 & 30.8435 & 0.260328 & -17.4693 \\
2.10 & 182.429 & 30.0052 & 0.161886 & -10.8634 \\
2.20 & 175.786 & 29.1768 & 0.0685114 & -4.59747 \\
2.30 & 169.328 & 28.3518 & -0.0193471 & 1.29829 \\
2.40 & 162.990 & 27.5199 & -0.101231 & 6.79314 \\
2.50 & 156.723 & 26.6735 & -0.176737 & 11.8600 \\
2.60 & 150.493 & 25.8071 & -0.245492 & 16.4738 \\
2.70 & 144.243 & 24.9121 & -0.307151 & 20.6115 \\
2.80 & 137.942 & 23.9843 & -0.361458 & 24.2557 \\
2.90 & 131.585 & 23.0232 & -0.408220 & 27.3937 \\
3.00 & 125.125 & 22.0226 & -0.447320 & 30.0175 \\
3.20 & 111.965 & 19.9190 & -0.502678 & 33.7323 \\
3.40 & 98.5935 & 17.7071 & -0.528954 & 35.4956 \\
3.60 & 85.2803 & 15.4469 & -0.529418 & 35.5267 \\
3.80 & 72.4156 & 13.2191 & -0.509111 & 34.1640 \\
4.00 & 60.3781 & 11.1023 & -0.473173 & 31.7524 \\
4.20 & 49.4911 & 9.16421 & -0.428052 & 28.7245 \\
4.40 & 39.9395 & 7.44650 & -0.378058 & 25.3697 \\
4.60 & 31.7964 & 5.96901 & -0.328370 & 22.0354 \\
4.80 & 25.0220 & 4.72985 & -0.279521 & 18.7573 \\
5.00 & 19.4962 & 3.71147 & -0.236301 & 15.8570 \\
5.20 & 15.0626 & 2.88852 & -0.197415 & 13.2476 \\
5.40 & 11.5579 & 2.23328 & -0.163798 & 10.9917 \\
5.60 & 8.81557 & 1.71698 & -0.134138 & 9.00133 \\
5.80 & 6.68912 & 1.31383 & -0.110457 & 7.41224 \\
6.00 & 5.05561 & 1.00182 & -0.0894048 & 5.99953 \\
\end{tabular}
\end{ruledtabular}
\end{table}
The data in Tab.~\ref{Tconstants} were interpolated at internuclear distances $R$ between 0 and 5 bohrs,
and extrapolated for $R>5$ bohrs by fitting $a_8/R^8+a_9/R^9+a_{10}/R^{10}+a_{11}/R^{11}$. 
This particular choice of powers of $R$ being in agreement with
numerical data, does not affect averaged results within five significant digits for the low lying levels. 
The averaged values, according to Eq. (\ref{Ehfsaver}), 
were evaluated with the nuclear wave function corresponding to a $(v,J)$ rovibrational level. 
This function is a solution of the radial nuclear equation, with nuclear masses and with the highly
accurate BO potential obtained in \cite{Pachucki:10b}, using the DVR method \cite{Colbert:92,Groenenboom:93}.
Numerical results for selected low lying states of HD are shown in Table III, while for an arbitrary rovibrational
level they can be obtained from the updated version of the publicly available H2Spectre computer code~\cite{H2Spectre}.

\renewcommand*{\arraystretch}{1.0}
 \begin{table*}[!ht] %\scriptsize
 \caption{Theoretically predicted hyperfine splitting parameters and levels (in kHz) for a selection of the lowest rovibrational levels $(v,J)$. The energy shifts $\delta E_F^\pm$ are labeled with the total angular momentum 
 $F$ and with $\pm$, which distinguishes between sublevels of the same $F$ but different energy.}
\label{THFS}
% \begin{ruledtabular}
 \begin{tabular}{c@{\extracolsep{\fill}}*{3}{x{5.4}}x{5.4}*{6}{x{6.3}}}
\hline\hline
$(v,J)$& \cent{\langle c_p\rangle} & \cent{\langle c_d\rangle} & \cent{\langle d_1\rangle} & \cent{\langle d_2\rangle} & 
%\cent{\delta E_{F=J+\frac{3}{2}}} & \multicolumn{2}{c}{$\delta E_{F=J+\frac{1}{2}}$} & \multicolumn{2}{c}{$\delta E_{F=J-\frac{1}{2}}$} & \cent{\delta E_{F=J-\frac{3}{2}}} \\
\cent{\delta E_{J+\frac{3}{2}}} & \cent{\delta E_{J+\frac{1}{2}}^-} & \cent{\delta E_{J+\frac{1}{2}}^+} & \cent{\delta E_{J-\frac{1}{2}}^-} & \cent{\delta E_{J-\frac{1}{2}}^+} & \cent{\delta E_{J-\frac{3}{2}}} \\
 \hline
(0,1)   & 85.675 & 13.132 & 17.773 & -22.459 &  -58.3 &   -1.9 &  54.1 & -117.0 & 187.5 & - \\
(0,2)   & 84.970 & 13.028 & 17.650 & -22.212 & -114.3 &  -30.1 &  67.7 & -115.2 & 209.8 & 155.1 \\
(0,3)   & 83.930 & 12.874 & 17.468 & -21.850 & -168.2 &  -67.3 &  90.8 & -135.9 & 244.4 & 210.6 \\
(0,4)   & 82.573 & 12.674 & 17.231 & -21.377 & -219.6 & -105.6 & 115.5 & -159.0 & 279.2 & 262.4 \\
(1,1)   & 84.067 & 12.846 & 17.225 & -22.305 &  -57.4 &   -1.6 &  53.8 & -115.8 & 183.7 & - \\
(1,2)   & 83.356 & 12.742 & 17.102 & -22.057 & -112.4 &  -29.0 &  66.8 & -113.5 & 206.2 & 150.9 \\
(1,3)   & 82.308 & 12.588 & 16.922 & -21.691 & -165.2 &  -65.4 &  89.3 & -133.6 & 240.1 & 205.4 \\
(1,4)   & 80.942 & 12.387 & 16.686 & -21.216 & -215.6 & -102.9 & 113.4 & -156.2 & 274.2 & 256.2 \\
(2,1)   & 82.183 & 12.524 & 16.654 & -22.043 &  -56.3 &   -1.4 &  53.3 & -114.1 & 179.3 & - \\
(2,2)   & 81.470 & 12.420 & 16.533 & -21.794 & -110.1 &  -27.8 &  65.5 & -111.4 & 201.8 & 146.3 \\
(2,3)   & 80.418 & 12.265 & 16.354 & -21.427 & -161.6 &  -63.3 &  87.5 & -130.9 & 235.0 & 199.8 \\
(2,4)   & 79.048 & 12.064 & 16.120 & -20.950 & -210.7 &  -99.9 & 111.0 & -152.9 & 268.3 & 249.3 \\
\hline\hline
\end{tabular}
% \end{ruledtabular}
 \end{table*}

\renewcommand{\arraystretch}{1.0}
 \begin{table}[!hb] %\scriptsize
 \caption{Comparison of our theoretically predicted hyperfine splitting parameters (in kHz) with the available 
 experimental \cite{Quinn:58} and theoretical \cite{Dupre:20} literature data.}
\label{THFSlit}
% \begin{ruledtabular}
 \begin{tabular}{c@{\extracolsep{\fill}}*{4}{x{5.5}}}
\hline\hline
$(v,J)$& \cent{\langle c_p\rangle}& \cent{\langle c_d\rangle}& \cent{\langle d_1\rangle}& \cent{\langle d_2\rangle}\\
 \hline
(0,1)     & 85.675(60)& 13.132(9) & 17.773(12)& -22.459(16)\\
Exper.$^a$& 85.600(18)& 13.122(11)& 17.761(12)& -22.454(6) \\
Theory$^b$&86.2832    & 13.2450   & 17.8317   & -22.66493  \\[2ex]
(1,1)     & 84.067(60)    & 12.846(9)    & 17.225(12)    & -22.305(16)    \\
Theory$^b$&85.0775    & 13.0599   & 17.2842   & -22.50968  \\[2ex]
(2,1)     & 82.183(60)    & 12.524(9)    & 16.654(12)    & -22.043(16)    \\
Theory$^b$&83.5670    & 12.8280   & 16.7190   & -22.25516  \\
\hline\hline
\end{tabular}
% \end{ruledtabular}
\flushleft
$^a${\footnotesize W. E. Quinn {\em et al.} (1958) \cite{Quinn:58}.}\\
$^b${\footnotesize P. Dupr\'e (2020) \cite{Dupre:20}.}
 \end{table} 

Considering the quadrupole moment of deuteron, it can be determined from the electric quadrupole 
coupling constant $d_2$, obtained from Ramsey measurements performed for HD in $J$=1 level \cite{Quinn:58}, 
and for D$_2$ in $J$=1 and $J$=2 levels \cite{Code:71} in the ground vibrational state.
Among them, the most accurate is the value
\begin{align}
  d_2=-22.5037(14)\,{\rm kHz}
\end{align}
obtained from the measurement for the $J$=1 level of D$_2$, which was later refined in \cite{Reid:73}.
Our value for the gradient of the electric field for this level is
\begin{align}
  \langle q \rangle = 0.33535(18)\text{ a.u.}
\end{align}
The quadrupole moment, obtained using this value and Eq.~(\ref{Ed2}), is
\begin{align}\label{EQD}
  Q = -\frac{d_2}{2\,Ry\,c}\,\frac{10\,\lambdabar^2}{\alpha^2\,\langle q\rangle} = 0.2856(2)\,{\rm fm}^2\,.
  \end{align}
Its uncertainty comes from the neglected nonadiabatic effects, which are of the order of
the ratio of the electron mass to the reduced nuclear mass $m_n(\text{D}_2)$. 
This quadrupole moment $Q$ is used in Tab.~\ref{Tconstants} to obtain the electric quadrupole constant $d_2$
as a function of $R$ and in Tab.~\ref{THFS} for various rovibrational levels.

A similar relative uncertainty of $1/m_n$(HD)$\,\approx 0.8\cdot 10^{-3}$ due to the omitted nonadiabatic effects
is assumed for all hyperfine constants in Tables~\ref{THFS} and~\ref{THFSlit}.
Because this uncertainty is larger than our numerical uncertainties, the latter were neglected.
Moreover, we expect that theoretical predictions for $d_2$ shall be in fact more accurate
due to partial cancellation between nonadiabatic effects in D$_2$ and HD.
Indeed, in comparison to measurements performed by Ramsey {\em et.al.} \cite{Quinn:58},
see Table \ref{THFSlit}, all our values differ by about $\sigma$, with the exception of $d_2$,
which differs by only $\sigma/3$. In conclusion, all our results are in agreement
with experimental values.

Considering the comparison with previous theoretical calculations,
our quadrupole moment of the deuteron $Q=0.2856(2)$ fm$^2$ differs within uncertainties from
values obtained by Pavanello {\em et al.} \cite{Pavanello:10} $0.285783(30)$ fm$^2$, 
Bishop and Cheung \cite{Bishop:79} $0.2862(15)$ fm$^2$, and
Reid and Vaida \cite{Reid:72,Reid:75Errata} $0.2860(15)$ fm$^2$. 
Surprisingly, the result of Ref.~\onlinecite{Pavanello:10} has tighter error bars than that of ours,
most probably due to underestimation of nonadiabatic effects. Moreover, 
results of the hyperfine parameters for the HD molecule, but without any uncertainties,
have recently been obtained by Dupr\'e \cite{Dupre:20},
who considered three vibrational levels ($v=0,\,1,\,2$) with the rotational quantum number $J=1$.
His results are presented in Tab.~\ref{THFSlit} after conversion from a different notation
($d_1=2\,c_\mathrm{dip}/5$, $d_2=-c_\mathrm{quad}/10$).
As one can notice, his results for the $v=0$, $J=1$ level differ from the experimental
ones by several hundreds of Hz. Similar difference appears in comparison with our values
and this difference grows with the vibrational quantum number.

\section{Hyperfine structure and individual transition rates}

The hyperfine structure for each molecular level $(v,J)$ is obtained by diagonalization of
the Hamiltonian $H_{\rm hfs}(v,J)$ in Eq.~(\ref{01}). We perform this diagonalization 
in the basis of $|J,M_J;I_p,M_p;I_d,M_d\rangle$ states because this basis is convenient
for the later calculation of transition rates. Explicit formulas for eigenvalues $\delta E_F$
($F$ is the total angular momentum) for $J=1,\ldots,4$
are given in Appendix, while their numerical values are presented in Table III.
These eigenvalues represent the shift of the molecular hyperfine level with respect to the centroid.
These hyperfine sublevels extend in the range of several hundreds of kHz, e.g. 300 kHz for the $(2,1)$ state
and 500 kHz for the $(0,4)$ state.
Regarding their uncertainty, it mainly comes from the neglected nonadiabatic effects, and this is already included
in the hyperfine coefficient. However, we do not perform detailed analysis of the resulting uncertainty
of individual hyperfine levels, but in general it should be about $0.1$ kHz, if not less.

Regarding hyperfine resolved transition rates, the main factor determining the line intensity
is the square of the transition electric dipole moment. Because we are interested here in relative intensities,
we consider only its angular part, which is
\begin{align}
%|\vec d_{if}|^2 &= \langle\chi_f|d(R)|\chi_i\rangle^2\,\sum_{M_i} \sum_{M_f} \langle F_f,M_f|\vec{n}| F_i,M_i\rangle^2\\
|\vec d_{if}|^2 &=\sum_{M_i} \sum_{M_f} |\langle F_f,M_f|\vec{n}| F_i,M_i\rangle|^2\,,
\end{align}
where the double sum goes over all the possible projections of the total angular momenta
of both the final and initial state.
The above matrix elements were evaluated with the eigenfunctions of the $H_{\rm hfs}(v,J)$
in the previously mentioned basis of $|J,M_J;I_p,M_p;I_d,M_d\rangle$ functions.

We now turn to analysis of recent measurements.
There are several very accurate measurements reported in literature concerning 
the infrared absorption in HD. All of them have uncertainties much below 100 kHz assigned
to the transition energy. We have determined the hyperfine splittings for the initial
and final states involved in these transitions and estimated the relative intensities
for all the hyperfine components. The obtained stick spectra were dressed with the Lorentzian
line shapes in order to simulate the overall line shape.

\subsubsection{$R_2(1)$ transition}

\begin{figure}
\includegraphics[scale=0.25]{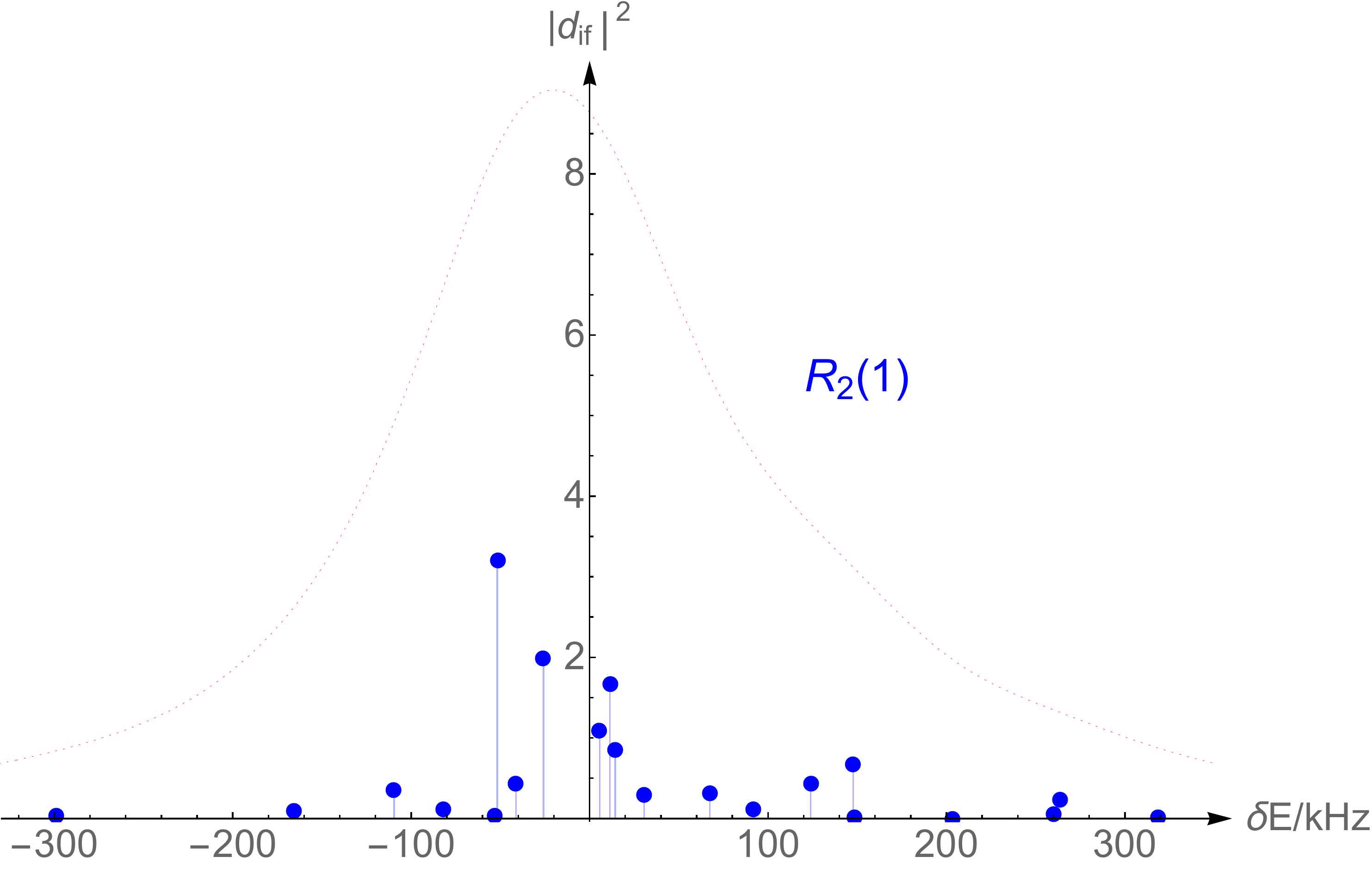}
\caption{Graphical representation of the $R_2(1)$ line.
Dotted line represents a Lorentzian line shape ($\omega=150$ kHz) superimposed on the stick spectrum.}
\label{FR1}
\end{figure}

The first transition line of interest is the $R_2(1)$ or $(0,1)\to(2,2)$ line.
This transition was studied by three different experimental groups reporting the following
transition energies: 
Fasci {\em et al.} \cite{Fasci:18}    $217\,105\,181.581(94)$ MHz,
Cozijn {\em et al.} \cite{Cozijn:18}  $217\,105\,181.895(20)$ MHz,
and Tao {\em et al.} \cite{Tao:18}        $217\,105\,182.79(3)(8)$ MHz.
The disagreement between these results can, at least partially, be attributed to the unresolved
hyperfine structure of the line. A thorough analysis of the pressure-dependent line shapes
related to the hyperfine splitting of the involved rovibrational levels has been performed 
in \cite{Diouf:19} and resulted in a refined transition frequency for this line equal to
$217\,105\,181.901(50)$ MHz. The corresponding theoretical prediction for this transition 
is $217\,105\,180.2(0.9)$ MHz \cite{CPKP:18}. Table~\ref{TR1} and Fig.~\ref{FR1} present
the theoretical hyperfine spectrum for this absorption line.

\renewcommand*{\arraystretch}{1.2}
\begin{table}[!h] %\scriptsize
\caption{Theoretically predicted line list of the hyperfine splitting of the $R_2(1)$ line.
$F$ is the total angular momentum quantum number. The label $+$ or $-$ distinguishes levels
of the same $F$ but different energy (see Tab.~\ref{THFS}).}
\label{TR1}
\begin{tabular*}{0.5\textwidth}{l@{\extracolsep{\fill}}x{5.2}x{2.3}}
\hline\hline
$\ \left|F_i\right\rangle\ \to\ \left|F_f\right\rangle$&  \cent{\delta E/\text{kHz}}  &  \cent{|\vec{d}_{if}|^2} \\
\hline
$\left|\frac{1}{2}+\right\rangle \to \left|\frac{3}{2}-\right\rangle$ & -298.9 & 0.042 \\
$\left|\frac{3}{2}+\right\rangle \to \left|\frac{3}{2}-\right\rangle$ & -165.5 & 0.101 \\
$\left|\frac{3}{2}-\right\rangle \to \left|\frac{3}{2}-\right\rangle$ & -109.5 & 0.344 \\
$\left|\frac{3}{2}+\right\rangle \to \left|\frac{5}{2}-\right\rangle$ & -82.0 & 0.116 \\
$\left|\frac{5}{2}\phantom{-}\right\rangle \to \left|\frac{3}{2}-\right\rangle$ & -53.1 & 0.024 \\
$\left|\frac{5}{2}\phantom{-}\right\rangle \to \left|\frac{7}{2}\phantom{-}\right\rangle$ & -51.8 & 3.200 \\
$\left|\frac{1}{2}+\right\rangle \to \left|\frac{1}{2}\phantom{-}\right\rangle$ & -41.2 & 0.435 \\
$\left|\frac{3}{2}-\right\rangle \to \left|\frac{5}{2}-\right\rangle$ & -25.9 & 1.987 \\
$\left|\frac{1}{2}-\right\rangle \to \left|\frac{3}{2}-\right\rangle$ & 5.6 & 1.089 \\
$\left|\frac{3}{2}+\right\rangle \to \left|\frac{5}{2}+\right\rangle$ & 11.4 & 1.659 \\
$\left|\frac{1}{2}+\right\rangle \to \left|\frac{3}{2}+\right\rangle$ & 14.3 & 0.857 \\
$\left|\frac{5}{2}\phantom{-}\right\rangle \to \left|\frac{5}{2}-\right\rangle$ & 30.5 & 0.296 \\
$\left|\frac{3}{2}-\right\rangle \to \left|\frac{5}{2}+\right\rangle$ & 67.5 & 0.317 \\
$\left|\frac{3}{2}+\right\rangle \to \left|\frac{1}{2}\phantom{-}\right\rangle$ & 92.2 & 0.116 \\
$\left|\frac{5}{2}\phantom{-}\right\rangle \to \left|\frac{5}{2}+\right\rangle$ & 123.9 & 0.424 \\
$\left|\frac{3}{2}+\right\rangle \to \left|\frac{3}{2}+\right\rangle$ & 147.7 & 0.674 \\
$\left|\frac{3}{2}-\right\rangle \to \left|\frac{1}{2}\phantom{-}\right\rangle$ & 148.2 & 0.018 \\
$\left|\frac{3}{2}-\right\rangle \to \left|\frac{3}{2}+\right\rangle$ & 203.7 & 0.000 \\
$\left|\frac{5}{2}\phantom{-}\right\rangle \to \left|\frac{3}{2}+\right\rangle$ & 260.1 & 0.056 \\
$\left|\frac{1}{2}-\right\rangle \to \left|\frac{1}{2}\phantom{-}\right\rangle$ & 263.3 & 0.232 \\
$\left|\frac{1}{2}-\right\rangle \to \left|\frac{3}{2}+\right\rangle$ & 318.8 & 0.013 \\
\hline\hline
\end{tabular*}
\end{table}

\subsubsection{$P_2(1)$ transition}

\begin{figure}[!hb]
\includegraphics[scale=0.25]{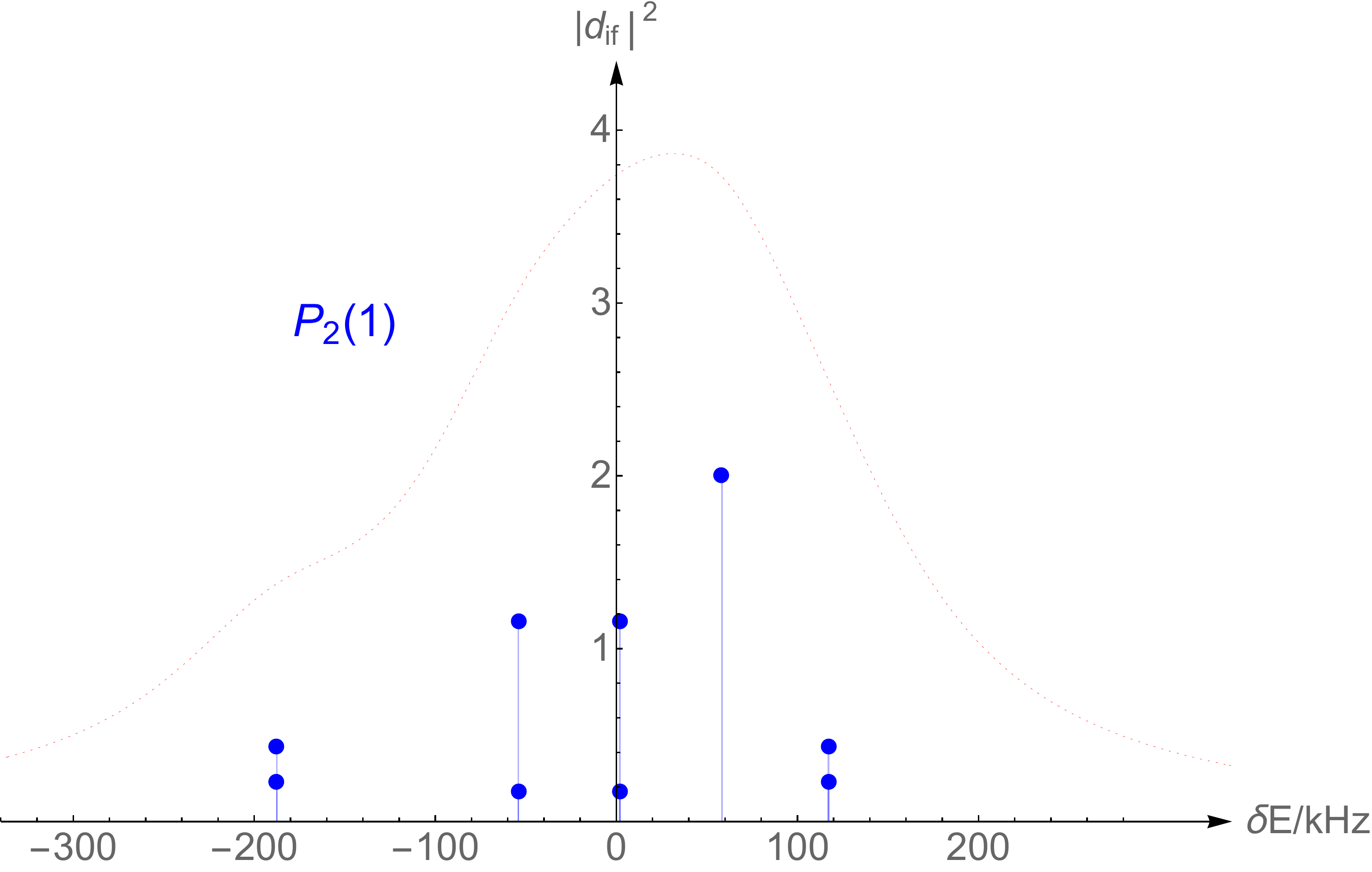}
\caption{Graphical representation of the $P_2(1)$ line.
Dotted line represents a Lorentzian line shape ($\omega=150$ kHz) superimposed on the stick spectrum.}
\label{FP1}
\end{figure}

Diouf {\em et al.} \cite{Diouf:20} measured the $P_2(1)$ or $(0,1)\to(2,0)$ absorption line 
and, employing the line shape analysis mentioned above, obtained the transition frequency 
$209\,784\,242\,007(20)$ kHz. The uncertainty of 20 kHz is more than an order of magnitude 
smaller than the extent of the hyperfine splitting (ca. 300 kHz). 
The calculated frequency for this transition line is $209\,784\,240.1(1.0)$ MHz \cite{H2Spectre,Komasa:19}.
The theoretical model of the hyperfine spectrum is shown in Table~\ref{TP1} and Figure~\ref{FP1}.

 \begin{table}[!h] %\scriptsize
 \caption{Theoretically predicted line list of the hyperfine splitting of the $P_2(1)$ line.
 $F$ is the total angular momentum quantum number. The label $+$ or $-$ distinguishes levels
of the same $F$ but different energy (see Tab.~\ref{THFS}).}
\label{TP1}
 \begin{tabular*}{0.5\textwidth}{l@{\extracolsep{\fill}}x{5.2}x{2.3}}
\hline\hline
$\ \left|F_i\right\rangle\ \to\ \left|F_f\right\rangle$&  \cent{\delta E/\text{kHz}}  &  \cent{|\vec{d}_{if}|^2} \\
\hline
$\left|\frac{1}{2}+\right\rangle \to \left|\frac{1}{2}-\right\rangle$ & -187.5 & 0.232 \\
$\left|\frac{1}{2}+\right\rangle \to \left|\frac{3}{2}\phantom{-}\right\rangle$ & -187.5 & 0.435 \\
$\left|\frac{3}{2}+\right\rangle \to \left|\frac{1}{2}-\right\rangle$ & -54.1 & 0.177 \\
$\left|\frac{3}{2}+\right\rangle \to \left|\frac{3}{2}\phantom{-}\right\rangle$ & -54.1 & 1.156 \\
$\left|\frac{3}{2}-\right\rangle \to \left|\frac{3}{2}\phantom{-}\right\rangle$ & 1.9 & 0.177 \\
$\left|\frac{3}{2}-\right\rangle \to \left|\frac{1}{2}-\right\rangle$ & 1.9 & 1.156 \\
$\left|\frac{5}{2}\phantom{-}\right\rangle \to \left|\frac{3}{2}\phantom{-}\right\rangle$ & 58.3 & 2.000 \\
$\left|\frac{1}{2}-\right\rangle \to \left|\frac{3}{2}\phantom{-}\right\rangle$ & 117.0 & 0.232 \\
$\left|\frac{1}{2}-\right\rangle \to \left|\frac{1}{2}-\right\rangle$ & 117.0 & 0.435 \\
\hline\hline
\end{tabular*}
\end{table}

\subsubsection{$R_1(0)$ transition}

\begin{figure}[!hb]
\includegraphics[scale=0.25]{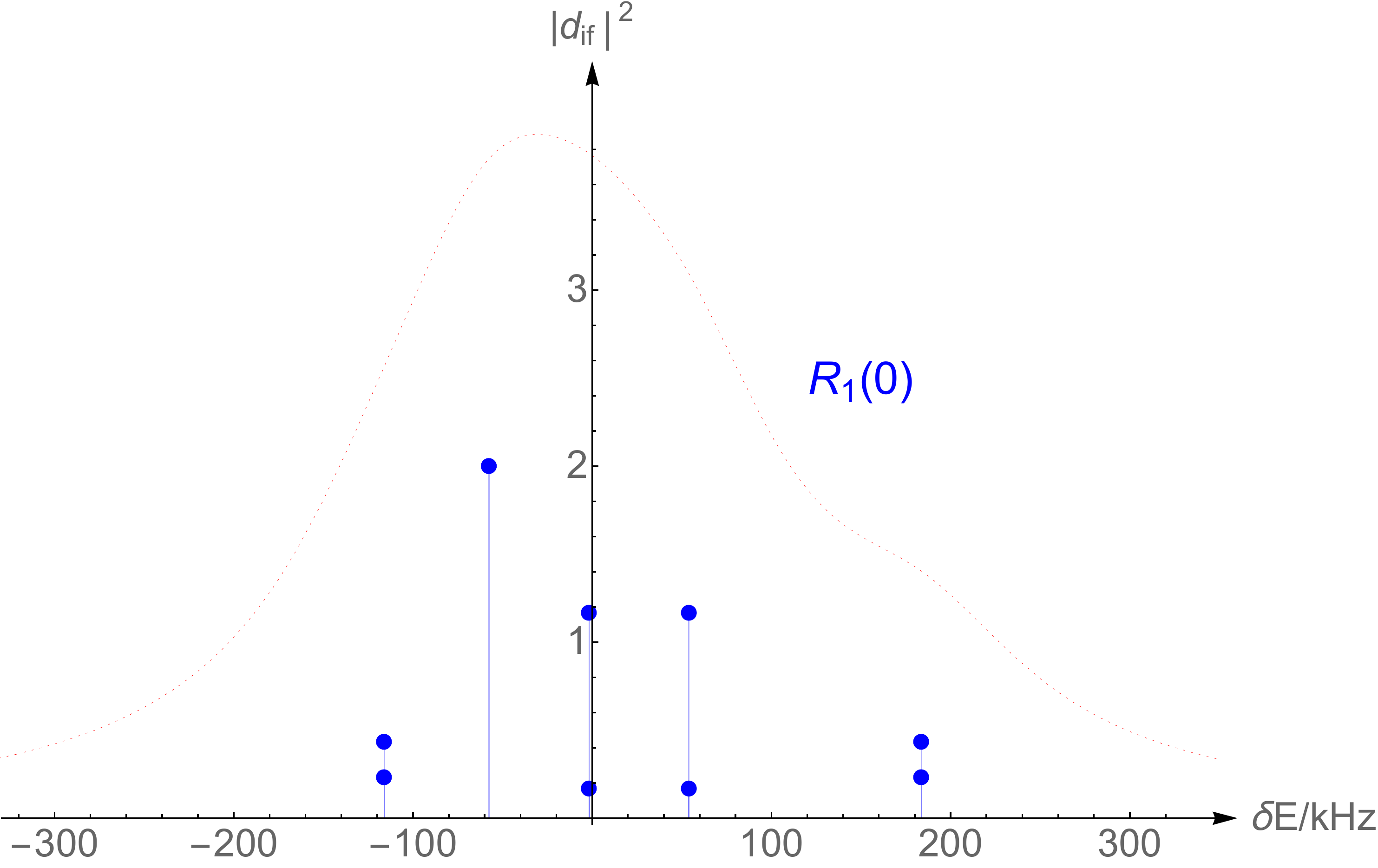}
\caption{Graphical representation of the $R_1(0)$ line.
Dotted line represents a Lorentzian line shape ($\omega=150$ kHz) superimposed on the stick spectrum.}
\label{FR0}
\end{figure}

Fast and Meek \cite{Fast:20} recently measured the $R_1(0)$ i.e. $(0,0)\to(1,1)$ transition using double resonance spectroscopy
in a molecular beam. The transition frequency of $111\,448\,815\,477(13)$ kHz was determined
with unprecedented relative accuracy of $1.2\cdot 10^{-10}$. The absolute uncertainty of 13 kHz is
over 20 times smaller than the 300 kHz extent of hyperfine splitting in the upper rovibrational level.
This experimental result can be compared with the theoretically predicted frequency of $111\,448\,814.5(6)$ MHz \cite{H2Spectre,Komasa:19}.
A theoretical absorption spectrum pertinent to this transition is shown in Table~\ref{TR0} and Figure~\ref{FR0}.

 \begin{table}[!h] %\scriptsize
 \caption{Theoretically predicted line list of the hyperfine splitting of the $R_1(0)$ line.
 $F$ is the total angular momentum quantum number. The label $+$ or $-$ distinguishes levels
of the same $F$ but different energy (see Tab.~\ref{THFS}).}
\label{TR0}
 \begin{tabular*}{0.5\textwidth}{l@{\extracolsep{\fill}}x{5.2}x{2.3}}
\hline\hline
$\ \left|F_i\right\rangle\ \to\ \left|F_f\right\rangle$&  \cent{\delta E/\text{kHz}}  &  \cent{|\vec{d}_{if}|^2} \\
\hline
$\left|\frac{3}{2}\phantom{-}\right\rangle \to \left|\frac{1}{2}-\right\rangle$ & -115.8 & 0.234 \\
$\left|\frac{1}{2}-\right\rangle \to \left|\frac{1}{2}-\right\rangle$ & -115.8 & 0.433 \\
$\left|\frac{3}{2}\phantom{-}\right\rangle \to \left|\frac{5}{2}\phantom{-}\right\rangle$ & -57.4 & 2.000 \\
$\left|\frac{3}{2}\phantom{-}\right\rangle \to \left|\frac{3}{2}-\right\rangle$ & -1.6 & 0.170 \\
$\left|\frac{1}{2}-\right\rangle \to \left|\frac{3}{2}-\right\rangle$ & -1.6 & 1.163 \\
$\left|\frac{1}{2}-\right\rangle \to \left|\frac{3}{2}+\right\rangle$ & 53.8 & 0.170 \\
$\left|\frac{3}{2}\phantom{-}\right\rangle \to \left|\frac{3}{2}+\right\rangle$ & 53.8 & 1.163 \\
$\left|\frac{1}{2}-\right\rangle \to \left|\frac{1}{2}+\right\rangle$ & 183.7 & 0.234 \\
$\left|\frac{3}{2}\phantom{-}\right\rangle \to \left|\frac{1}{2}+\right\rangle$ & 183.7 & 0.433 \\
\hline\hline
\end{tabular*}
\end{table}

\section{Summary and outlook}
We performed the derivation and the numerical calculation of the leading hyperfine interactions
in the HD molecule. Moreover, we obtained hyperfine constants for all low lying levels of HD
and compared with experimental and previous theoretical results.
The accuracy of our calculations is limited by the unknown nonadiabatic effects, which are estimated by
the ratio of the electron mass to the reduced nuclear mass. Very good agreement is achieved
with results of the measurements by Ramsey \cite{Quinn:58} for the first rotational state of HD.
From the measurement of the $d_2$ constant in the D$_2$ molecule \cite{Code:71,Reid:73} we determined
the value of the deuteron quadrupole moment in agreement with
the previous determinations, but with greater accuracy.
However, our results for the hyperfine constants in HD differ from the previous calculations in Ref. \onlinecite{Dupre:20}
by hundreds of Hz for $v=0$ and this difference grows with the vibrational number.

All the nonadiabatic effects, which presently limit our accuracy, can be calculated with the use 
of a very accurate nonadiabatic wave function expanded in explicitly correlated exponential \cite{PK:18b} basis.
This requires, however, the development of integrals with quadratic inverse powers of interparticle distances,
and we are presently pursuing this project.

Although we did not calculate relativistic corrections to the hyperfine coefficients, we stress
their importance in achieving high-precision theoretical predictions for the
molecular hfs. They are of particular interest for
an improved determination of the deuteron quadrupole moment.
These relativistic corrections can be calculated in the BO approximation, 
as previously done for the nuclear spin-spin coupling \cite{PKP:18}.
To perform such calculations, however, appropriate formulas have to be derived.
Ramsey in 1953 \cite{Ramsey:53}, worked out formulas for the nuclear spin-spin interactions. 
In a similar way, one can obtain relativistic corrections to the electric quadrupole moment
and to the spin-rotation constants. Having a pertinent theoretical framework, one can calculate
all these hyperfine constants with a relative accuracy of $\alpha^3/\pi$, limited by
the unknown QED effects. Numerically it is about $10^{-7}$, and we claim that this accuracy
can be achieved for all the hyperfine parameters  in HD, H$_2$ and D$_2$ molecules.
Such accuracy will give an opportunity for high precision tests of molecular hyperfine interactions,
provided that measurements of similar accuracy are performed.
We hope that the present work will encourage experimentalists to undertake this challenge.

\acknowledgments
This research was supported by National Science Center (Poland) Grant No. 2016/23/B/ST4/01821 as well as by a computing grant from Pozna\'n Supercomputing and Networking Center and by PL-Grid Infrastructure.

\appendix
\section{Eigenvalues of $H_{\rm hfs}$ for the lowest $J$}

In this section analytic formulas for the eigenvalues of the hyperfine Hamiltonian are provided.
The labeling of the eigenvalues $\delta E$ corresponds to that used in Table~\ref{THFS}.
For simplicity, the symbol of the rovibrational averaging was dropped here, i.e. $c_p\equiv \langle c_p\rangle$ etc.

\subsection*{$J=1$}

\vspace{-5ex}
\begin{align}
\delta E_{5/2}&=-c_d-\frac{c_p}{2}+\frac{d_1}{2}+\frac{d_2}{2}\\
\delta E_{3/2}^\pm&=\frac{c_p}{4}-d_1-d_2\pm\frac{1}{4} \sqrt{A_1}\\
\delta E_{1/2}^\pm&=\frac{3 c_d}{2}+\frac{c_p}{4}+\frac{5 d_1}{4}+\frac{5 d_2}{4}\pm\frac{1}{4} \sqrt{B_1}\\
A_1&=-16 c_d c_p-16 d_1 c_p+24 d_2 c_p+16 c_d^2-8 d_1 c_d\nonumber\\
		&\quad-48 d_2 c_d+9 c_p^2+21 d_1^2+36 d_2^2+12 d_1 d_2\\
B_1&=-4 c_d c_p+50 d_1 c_p-30 d_2 c_p+4 c_d^2-20 d_1 c_d\nonumber\\
		&\quad+60 d_2 c_d+9 c_p^2+75 d_1^2+225 d_2^2-150 d_1 d_2
\end{align}

\subsection*{$J=2$}
\vspace{-5ex}
\begin{align}
\delta E_{7/2}&=-2 c_d-c_p+\frac{5 d_1}{7}+\frac{5 d_2}{7}\\
\delta E_{5/2}^\pm&=-\frac{c_d}{2}+\frac{c_p}{4}-\frac{25 d_1}{28}-\frac{25
   d_2}{28}\pm\frac{1}{28} \sqrt{B_2}\\
\delta E_{3/2}^\pm&=2 c_d+\frac{c_p}{4}\pm\frac{1}{4} \sqrt{A_2}\\
\delta E_{1/2}&=3 c_d+\frac{3 c_p}{2}+\frac{5 d_1}{2}+\frac{5 d_2}{2}\\
A_2&=-32 c_d c_p+40 d_1 c_p-80 d_2 c_p+16 c_d^2\nonumber\\
   &\quad-40 d_1 c_d+80 d_2 c_d+25 c_p^2+25 d_1^2\nonumber\\
   &\quad+100 d_2^2-100 d_1 d_2\\
B_2&=-2548 c_d c_p-1190 d_1 c_p+2730 d_2 c_p\nonumber\\
   &\quad+1764 c_d^2+140 d_1 c_d-3780 d_2 c_d+1225 c_p^2\nonumber\\
   &\quad+975 d_1^2+2025 d_2^2-150 d_1 d_2
\end{align}

\subsection*{$J=3$}
\vspace{-5ex}
\begin{align}
\delta E_{9/2}&=-3 c_d-\frac{3 c_p}{2}+\frac{5 d_1}{6}+\frac{5 d_2}{6}\\
\delta E_{7/2}^\pm&=-c_d+\frac{c_p}{4}-\frac{5 d_1}{6}-\frac{5 d_2}{6}\pm\frac{1}{12} \sqrt{B_3}\\
\delta E_{5/2}^\pm&=\frac{5 c_d}{2}+\frac{c_p}{4}-\frac{d_1}{4}-\frac{d_2}{4}\pm\frac{1}{4} \sqrt{A_3}\\
\delta E_{3/2}&=4 c_d+2 c_p+2 d_1+2 d_2\\
A_3&=-76 c_d c_p+46 d_1 c_p-114 d_2 c_p+36 c_d^2\nonumber\\
   &\quad-52 d_1 c_d+108 d_2 c_d+49 c_p^2+21 d_1^2\nonumber\\
   &\quad+81 d_2^2-78 d_1 d_2\\
B_3&=-936 c_d c_p-300 d_1 c_p+780 d_2 c_p+576 c_d^2\nonumber\\
   &\quad+120 d_1 c_d-960 d_2 c_d+441 c_p^2+175 d_1^2\nonumber\\
   &\quad+400 d_2^2-100 d_1 d_2
\end{align}

\subsection*{$J=4$}
\vspace{-5ex}
\begin{align}
\delta E_{11/2}&=-4 c_d-2 c_p+\frac{10 d_1}{11}+\frac{10 d_2}{11}\\
\delta E_{9/2}^\pm&=-\frac{3 c_d}{2}+\frac{c_p}{4}-\frac{35 d_1}{44}-\frac{35 d_2}{44}\pm\frac{1}{44} \sqrt{B_4}\\
\delta E_{7/2}^\pm&=3 c_d+\frac{c_p}{4}-\frac{5 d_1}{14}-\frac{5 d_2}{14}\pm\frac{1}{28} \sqrt{A_4}\\
\delta E_{5/2}&=5 c_d+\frac{5 c_p}{2}+\frac{25 d_1}{14}+\frac{25 d_2}{14}\\
A_4&=-6664 c_d c_p+2660 d_1 c_p-7140 d_2 c_p\nonumber\\
   &\quad+3136 c_d^2-3080 d_1 c_d+6720 d_2 c_d\nonumber\\
   &\quad+3969 c_p^2 +975 d_1^2+3600 d_2^2-3300 d_1 d_2\\
B_4&=-20812 c_d c_p-5170 d_1 c_p+14190 d_2 c_p\nonumber\\
   &\quad+12100 c_d^2+2860 d_1 c_d-16500 d_2 c_d\nonumber\\
   &\quad+9801 c_p^2+2325 d_1^2+5625 d_2^2-1950 d_1 d_2
\end{align}


\begin{thebibliography}{28}
\expandafter\ifx\csname natexlab\endcsname\relax\def\natexlab#1{#1}\fi
\expandafter\ifx\csname bibnamefont\endcsname\relax
  \def\bibnamefont#1{#1}\fi
\expandafter\ifx\csname bibfnamefont\endcsname\relax
  \def\bibfnamefont#1{#1}\fi
\expandafter\ifx\csname citenamefont\endcsname\relax
  \def\citenamefont#1{#1}\fi
\expandafter\ifx\csname url\endcsname\relax
  \def\url#1{\texttt{#1}}\fi
\expandafter\ifx\csname urlprefix\endcsname\relax\def\urlprefix{URL }\fi
\providecommand{\bibinfo}[2]{#2}
\providecommand{\eprint}[2][]{\url{#2}}

\bibitem[{\citenamefont{Quinn et~al.}(1958)\citenamefont{Quinn, Baker,
  LaTourrette, and Ramsey}}]{Quinn:58}
\bibinfo{author}{\bibfnamefont{W.~E.} \bibnamefont{Quinn}},
  \bibinfo{author}{\bibfnamefont{J.~M.} \bibnamefont{Baker}},
  \bibinfo{author}{\bibfnamefont{J.~T.} \bibnamefont{LaTourrette}},
  \bibnamefont{and} \bibinfo{author}{\bibfnamefont{N.~F.}
  \bibnamefont{Ramsey}}, \bibinfo{journal}{Phys. Rev.}
  \textbf{\bibinfo{volume}{112}}, \bibinfo{pages}{1929} (\bibinfo{year}{1958}).

\bibitem[{\citenamefont{Code and Ramsey}(1971)}]{Code:71}
\bibinfo{author}{\bibfnamefont{R.~F.} \bibnamefont{Code}} \bibnamefont{and}
  \bibinfo{author}{\bibfnamefont{N.~F.} \bibnamefont{Ramsey}},
  \bibinfo{journal}{Phys. Rev. A} \textbf{\bibinfo{volume}{4}},
  \bibinfo{pages}{1945} (\bibinfo{year}{1971}).

\bibitem[{\citenamefont{Fasci et~al.}(2018)\citenamefont{Fasci, Castrillo,
  Dinesan, Gravina, Moretti, and Gianfrani}}]{Fasci:18}
\bibinfo{author}{\bibfnamefont{E.}~\bibnamefont{Fasci}},
  \bibinfo{author}{\bibfnamefont{A.}~\bibnamefont{Castrillo}},
  \bibinfo{author}{\bibfnamefont{H.}~\bibnamefont{Dinesan}},
  \bibinfo{author}{\bibfnamefont{S.}~\bibnamefont{Gravina}},
  \bibinfo{author}{\bibfnamefont{L.}~\bibnamefont{Moretti}}, \bibnamefont{and}
  \bibinfo{author}{\bibfnamefont{L.}~\bibnamefont{Gianfrani}},
  \bibinfo{journal}{Phys. Rev. A} \textbf{\bibinfo{volume}{98}},
  \bibinfo{pages}{022516} (\bibinfo{year}{2018}).

\bibitem[{\citenamefont{Cozijn et~al.}(2018)\citenamefont{Cozijn, Dupr{\'e},
  Salumbides, Eikema, and Ubachs}}]{Cozijn:18}
\bibinfo{author}{\bibfnamefont{F.~M.~J.} \bibnamefont{Cozijn}},
  \bibinfo{author}{\bibfnamefont{P.}~\bibnamefont{Dupr{\'e}}},
  \bibinfo{author}{\bibfnamefont{E.~J.} \bibnamefont{Salumbides}},
  \bibinfo{author}{\bibfnamefont{K.~S.~E.} \bibnamefont{Eikema}},
  \bibnamefont{and} \bibinfo{author}{\bibfnamefont{W.}~\bibnamefont{Ubachs}},
  \bibinfo{journal}{Phys. Rev. Lett.} \textbf{\bibinfo{volume}{120}},
  \bibinfo{pages}{153002} (\bibinfo{year}{2018}).

\bibitem[{\citenamefont{Tao et~al.}(2018)\citenamefont{Tao, Liu, Pachucki,
  Komasa, Sun, Wang, and Hu}}]{Tao:18}
\bibinfo{author}{\bibfnamefont{L.-G.} \bibnamefont{Tao}},
  \bibinfo{author}{\bibfnamefont{A.-W.} \bibnamefont{Liu}},
  \bibinfo{author}{\bibfnamefont{K.}~\bibnamefont{Pachucki}},
  \bibinfo{author}{\bibfnamefont{J.}~\bibnamefont{Komasa}},
  \bibinfo{author}{\bibfnamefont{Y.~R.} \bibnamefont{Sun}},
  \bibinfo{author}{\bibfnamefont{J.}~\bibnamefont{Wang}}, \bibnamefont{and}
  \bibinfo{author}{\bibfnamefont{S.-M.} \bibnamefont{Hu}},
  \bibinfo{journal}{Phys. Rev. Lett.} \textbf{\bibinfo{volume}{120}},
  \bibinfo{pages}{153001} (\bibinfo{year}{2018}).

\bibitem[{\citenamefont{Dupr\'e}(2020)}]{Dupre:20}
\bibinfo{author}{\bibfnamefont{P.}~\bibnamefont{Dupr\'e}},
  \bibinfo{journal}{Phys. Rev. A} \textbf{\bibinfo{volume}{101}},
  \bibinfo{pages}{022504} (\bibinfo{year}{2020}).

\bibitem[{\citenamefont{{H2SPECTRE ver. 7.1 F}ortran~source
  code}(2020)}]{H2Spectre}
\bibinfo{author}{\bibnamefont{{H2SPECTRE ver. 7.1 F}ortran~source code}}
  (\bibinfo{year}{2020}), \bibinfo{note}{p. Czachorowski, Ph.D. thesis,
  University of Warsaw, Poland, 2019.},
  \urlprefix\url{https://www.fuw.edu.pl/~krp/codes.html;
  http://qcg.home.amu.edu.pl/qcg/public_html/H2Spectre.html;}.

\bibitem[{\citenamefont{Ramsey and Lewis}(1957)}]{Ramsey:57}
\bibinfo{author}{\bibfnamefont{N.~F.} \bibnamefont{Ramsey}} \bibnamefont{and}
  \bibinfo{author}{\bibfnamefont{H.~R.} \bibnamefont{Lewis}},
  \bibinfo{journal}{Phys. Rev.} \textbf{\bibinfo{volume}{108}},
  \bibinfo{pages}{1246} (\bibinfo{year}{1957}).

\bibitem[{\citenamefont{Ramsey}(1956)}]{Ramsey:56}
\bibinfo{author}{\bibfnamefont{N.}~\bibnamefont{Ramsey}},
  \emph{\bibinfo{title}{Molecular beams}} (\bibinfo{publisher}{Oxford Univ.
  Press}, \bibinfo{year}{1956}).

\bibitem[{\citenamefont{Pachucki}(2010{\natexlab{a}})}]{Pachucki:10}
\bibinfo{author}{\bibfnamefont{K.}~\bibnamefont{Pachucki}},
  \bibinfo{journal}{Phys. Rev. A} \textbf{\bibinfo{volume}{81}},
  \bibinfo{pages}{032505} (\bibinfo{year}{2010}{\natexlab{a}}).

\bibitem[{\citenamefont{{2018 CODATA recommended values}}(2018)}]{CODATA:18}
\bibinfo{author}{\bibnamefont{{2018 CODATA recommended values}}}
  (\bibinfo{year}{2018}),
  \urlprefix\url{https://physics.nist.gov/cuu/Constants}.

\bibitem[{\citenamefont{Reid and Chu}(1974)}]{Reid:74}
\bibinfo{author}{\bibfnamefont{R.~V.} \bibnamefont{Reid}} \bibnamefont{and}
  \bibinfo{author}{\bibfnamefont{A.~H.-M.} \bibnamefont{Chu}},
  \bibinfo{journal}{Phys. Rev. A} \textbf{\bibinfo{volume}{9}},
  \bibinfo{pages}{609} (\bibinfo{year}{1974}).

\bibitem[{\citenamefont{Pavanello et~al.}(2010)\citenamefont{Pavanello, Tung,
  and Adamowicz}}]{Pavanello:10}
\bibinfo{author}{\bibfnamefont{M.}~\bibnamefont{Pavanello}},
  \bibinfo{author}{\bibfnamefont{W.-C.} \bibnamefont{Tung}}, \bibnamefont{and}
  \bibinfo{author}{\bibfnamefont{L.}~\bibnamefont{Adamowicz}},
  \bibinfo{journal}{Phys. Rev. A} \textbf{\bibinfo{volume}{81}},
  \bibinfo{pages}{042526} (\bibinfo{year}{2010}).

\bibitem[{\citenamefont{Reid and Vaida}(1973)}]{Reid:73}
\bibinfo{author}{\bibfnamefont{R.~V.} \bibnamefont{Reid}} \bibnamefont{and}
  \bibinfo{author}{\bibfnamefont{M.~L.} \bibnamefont{Vaida}},
  \bibinfo{journal}{Phys. Rev. A} \textbf{\bibinfo{volume}{7}},
  \bibinfo{pages}{1841} (\bibinfo{year}{1973}).

\bibitem[{\citenamefont{Pachucki}(2010{\natexlab{b}})}]{Pachucki:10b}
\bibinfo{author}{\bibfnamefont{K.}~\bibnamefont{Pachucki}},
  \bibinfo{journal}{Phys. Rev. A} \textbf{\bibinfo{volume}{82}},
  \bibinfo{pages}{032509} (\bibinfo{year}{2010}{\natexlab{b}}).

\bibitem[{\citenamefont{Colbert and Miller}(1992)}]{Colbert:92}
\bibinfo{author}{\bibfnamefont{D.~T.} \bibnamefont{Colbert}} \bibnamefont{and}
  \bibinfo{author}{\bibfnamefont{W.~H.} \bibnamefont{Miller}},
  \bibinfo{journal}{The Journal of Chemical Physics}
  \textbf{\bibinfo{volume}{96}}, \bibinfo{pages}{1982} (\bibinfo{year}{1992}).

\bibitem[{\citenamefont{Groenenboom and Colbert}({1993})}]{Groenenboom:93}
\bibinfo{author}{\bibfnamefont{G.~C.} \bibnamefont{Groenenboom}}
  \bibnamefont{and} \bibinfo{author}{\bibfnamefont{D.~T.}
  \bibnamefont{Colbert}}, \bibinfo{journal}{J. Comp. Phys.}
  \textbf{\bibinfo{volume}{{99}}}, \bibinfo{pages}{9681}
  (\bibinfo{year}{{1993}}).

\bibitem[{\citenamefont{Bishop and Cheung}(1979)}]{Bishop:79}
\bibinfo{author}{\bibfnamefont{D.~M.} \bibnamefont{Bishop}} \bibnamefont{and}
  \bibinfo{author}{\bibfnamefont{L.~M.} \bibnamefont{Cheung}},
  \bibinfo{journal}{Phys. Rev. A} \textbf{\bibinfo{volume}{20}},
  \bibinfo{pages}{381} (\bibinfo{year}{1979}).

\bibitem[{\citenamefont{Reid and Vaida}(1972)}]{Reid:72}
\bibinfo{author}{\bibfnamefont{R.~V.} \bibnamefont{Reid}} \bibnamefont{and}
  \bibinfo{author}{\bibfnamefont{M.~L.} \bibnamefont{Vaida}},
  \bibinfo{journal}{Phys. Rev. Lett.} \textbf{\bibinfo{volume}{29}},
  \bibinfo{pages}{494} (\bibinfo{year}{1972}).

\bibitem[{\citenamefont{Reid and Vaida}(1975)}]{Reid:75Errata}
\bibinfo{author}{\bibfnamefont{R.~V.} \bibnamefont{Reid}} \bibnamefont{and}
  \bibinfo{author}{\bibfnamefont{M.~L.} \bibnamefont{Vaida}},
  \bibinfo{journal}{Phys. Rev. Lett.} \textbf{\bibinfo{volume}{34}},
  \bibinfo{pages}{1064} (\bibinfo{year}{1975}).

\bibitem[{\citenamefont{Diouf et~al.}(2019)\citenamefont{Diouf, Cozijn,
  Darqui\'{e}, Salumbides, and Ubachs}}]{Diouf:19}
\bibinfo{author}{\bibfnamefont{M.~L.} \bibnamefont{Diouf}},
  \bibinfo{author}{\bibfnamefont{F.~M.~J.} \bibnamefont{Cozijn}},
  \bibinfo{author}{\bibfnamefont{B.}~\bibnamefont{Darqui\'{e}}},
  \bibinfo{author}{\bibfnamefont{E.~J.} \bibnamefont{Salumbides}},
  \bibnamefont{and} \bibinfo{author}{\bibfnamefont{W.}~\bibnamefont{Ubachs}},
  \bibinfo{journal}{Opt. Lett.} \textbf{\bibinfo{volume}{44}},
  \bibinfo{pages}{4733} (\bibinfo{year}{2019}).

\bibitem[{\citenamefont{Czachorowski et~al.}(2018)\citenamefont{Czachorowski,
  Puchalski, Komasa, and Pachucki}}]{CPKP:18}
\bibinfo{author}{\bibfnamefont{P.}~\bibnamefont{Czachorowski}},
  \bibinfo{author}{\bibfnamefont{M.}~\bibnamefont{Puchalski}},
  \bibinfo{author}{\bibfnamefont{J.}~\bibnamefont{Komasa}}, \bibnamefont{and}
  \bibinfo{author}{\bibfnamefont{K.}~\bibnamefont{Pachucki}},
  \bibinfo{journal}{Phys. Rev. A} \textbf{\bibinfo{volume}{98}},
  \bibinfo{pages}{052506} (\bibinfo{year}{2018}).

\bibitem[{\citenamefont{Diouf et~al.}(2020)\citenamefont{Diouf, Cozijn, Lai,
  Salumbides, and Ubachs}}]{Diouf:20}
\bibinfo{author}{\bibfnamefont{M.~L.} \bibnamefont{Diouf}},
  \bibinfo{author}{\bibfnamefont{F.~M.~J.} \bibnamefont{Cozijn}},
  \bibinfo{author}{\bibfnamefont{K.-F.} \bibnamefont{Lai}},
  \bibinfo{author}{\bibfnamefont{E.~J.} \bibnamefont{Salumbides}},
  \bibnamefont{and} \bibinfo{author}{\bibfnamefont{W.}~\bibnamefont{Ubachs}}
  (\bibinfo{year}{2020}), \bibinfo{note}{unpublished}.

\bibitem[{\citenamefont{Komasa et~al.}(2019)\citenamefont{Komasa, Puchalski,
  Czachorowski, \L{}ach, and Pachucki}}]{Komasa:19}
\bibinfo{author}{\bibfnamefont{J.}~\bibnamefont{Komasa}},
  \bibinfo{author}{\bibfnamefont{M.}~\bibnamefont{Puchalski}},
  \bibinfo{author}{\bibfnamefont{P.}~\bibnamefont{Czachorowski}},
  \bibinfo{author}{\bibfnamefont{G.}~\bibnamefont{\L{}ach}}, \bibnamefont{and}
  \bibinfo{author}{\bibfnamefont{K.}~\bibnamefont{Pachucki}},
  \bibinfo{journal}{Phys. Rev. A} \textbf{\bibinfo{volume}{100}},
  \bibinfo{pages}{032519} (\bibinfo{year}{2019}).

\bibitem[{\citenamefont{Fast and Meek}(2020)}]{Fast:20}
\bibinfo{author}{\bibfnamefont{A.}~\bibnamefont{Fast}} \bibnamefont{and}
  \bibinfo{author}{\bibfnamefont{S.~A.} \bibnamefont{Meek}}
  (\bibinfo{year}{2020}), \eprint{arXiv:2002.09333}.

\bibitem[{\citenamefont{Pachucki and Komasa}(2018)}]{PK:18b}
\bibinfo{author}{\bibfnamefont{K.}~\bibnamefont{Pachucki}} \bibnamefont{and}
  \bibinfo{author}{\bibfnamefont{J.}~\bibnamefont{Komasa}},
  \bibinfo{journal}{Phys. Chem. Chem. Phys.} \textbf{\bibinfo{volume}{20}},
  \bibinfo{pages}{26297} (\bibinfo{year}{2018}).

\bibitem[{\citenamefont{Puchalski et~al.}(2018)\citenamefont{Puchalski, Komasa,
  and Pachucki}}]{PKP:18}
\bibinfo{author}{\bibfnamefont{M.}~\bibnamefont{Puchalski}},
  \bibinfo{author}{\bibfnamefont{J.}~\bibnamefont{Komasa}}, \bibnamefont{and}
  \bibinfo{author}{\bibfnamefont{K.}~\bibnamefont{Pachucki}},
  \bibinfo{journal}{Phys. Rev. Lett.} \textbf{\bibinfo{volume}{120}},
  \bibinfo{pages}{083001} (\bibinfo{year}{2018}).

\bibitem[{\citenamefont{Ramsey}(1953)}]{Ramsey:53}
\bibinfo{author}{\bibfnamefont{N.~F.} \bibnamefont{Ramsey}},
  \bibinfo{journal}{Phys. Rev.} \textbf{\bibinfo{volume}{91}},
  \bibinfo{pages}{303} (\bibinfo{year}{1953}).

\end{thebibliography}
\end{document}